\documentclass[twocolumn,aps, groupedaddress,epsf]{revtex4}

\usepackage{graphicx}

\begin{document}

\title{Fabrication and Electric Field Dependent Transport Measurements of Mesoscopic Graphite Devices}

\author{Yuanbo Zhang}
\author{Joshua P. Small}
\author{William V. Pontius}
\author{Philip Kim}

\affiliation{Department of Physics and the Columbia Nanoscale Science
and Engineering Center, Columbia University, New York, New York 10027}

\begin{abstract}
We have developed a unique micromechanical method to extract
extremely thin graphite samples. Graphite crystallites
with thicknesses ranging from 10~-~100~nm and lateral size $\sim$~2~$\mu$m
are extracted from bulk. Mesoscopic graphite devices are
fabricated from these samples
for electric field dependent conductance measurements.
Strong conductance modulation as a function of gate voltage is
observed in the thinner crystallite devices. The temperature
dependent resistivity measurements show more boundary scattering
contribution in the thinner graphite samples.

\end{abstract}

\maketitle


The recent focus of research in graphitic materials, such as
fullerenes~\cite{Dresselhaus_Book_1996} and carbon
nanotubes~\cite{Dresselhaus_Book_1996, Dresselhaus_Book_2001} has
renewed interest in electronic transport in graphene, a single
atomic layer of graphite. While graphite is a semimetal with
strong electron and hole compensation, graphene is expected to be
a zero-gap semiconductor with a vanishing density of states at the
charge neutral point. Electrical transport in thin graphite
crystals composed of only a few graphene layers is of particular
interest in elucidating the evolution of electronic structure from
bulk single crystals to two-dimensional planar systems.
Experimental work to synthesize very thin graphitic layers
directly on top of a substrate \cite{Chem} or to extract using
chemical \cite{Viculis_2003}, or mechanical \cite{Lu&Ruoff_1999,
Ebbensen&Hiura_1995} pathways has been demonstrated to produce
graphitic samples with thicknesses ranging from 1~-~100 nm.
Systematic transport measurements have been carried out on
mesoscopic graphitic disks \cite{Dujardin&Ebbesen_2001} and
cleaved bulk crystals \cite{Ohashi_1997+} with sample thicknesses
approaching $\sim$~20~nm, exhibiting persisting bulk graphite
properties at these length scales. However, electric field
dependent charge transport has not been measured, due to the fact
that the size of the samples and the geometry of the devices are
inadequate for such a measurement.

In this letter, we present a unique method to extract mesoscopic
graphite crystallites, with thicknesses ranging from 10~-~100~nm,
from bulk highly oriented pyrolytic graphite
(HOPG) using micromechanical manipulation and microfabrication techniques.
Mesoscopic devices consisting of these graphite crystallites
are fabricated for electrical transport measurement.
A degenerately doped Si substrate serves as a back gate, allowing us
to probe the electrical properties as a function of charge concentration.
Electrical conductance of these mesoscopic
graphite devices exhibits significant gate voltage dependance.
From the temperature dependent resistivity measurements
in these samples, we deduce
the scattering rate of mesoscopic graphite of varying thicknesses.

\begin{figure}
\includegraphics[width=80mm]{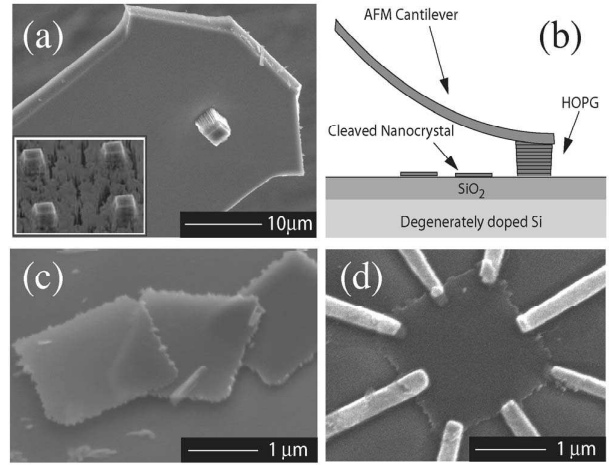}
\caption{(a) scanning electron microscope image of an HOPG
crystallite mounted on a microcantilever. Inset: Bulk HOPG
surface patterned by masked anisotropic
oxygen plasma etching. (b) Schematic drawing
of the micro-cleaving process. (c) Thin graphite samples
cleaved onto the SiO${_2}$/Si substrate. (d) A typical mesoscopic
device fabricated from a cleaved graphite sample.}
\label{fig1_cleaving}
\end{figure}

The graphite samples used in this study are extracted from
HOPG (\mbox{Grade-1}, Structure Probe, Inc.). Following a similar technique
demonstrated in
\cite{Lu&Ruoff_1999}, arrays of graphite micro-pillars
($\sim2\times2\times5$~$\mu$m$^3$)
are fabricated on the HOPG surface using micropatterning followed by
masked anisotropic oxygen plasma etching (Fig.~\ref{fig1_cleaving}(a) inset).
Once the array is formed, an
individual block of the pillar is removed from the
surface using a precision micro-manipulator under a high
resolution optical microscope. The detached graphite block is
then transferred to a micro-machined silicon cantilever (MikroMasch),
where it is glued down by a small amount of ultra
violet sensitive epoxy (Fig. \ref{fig1_cleaving}(a)).

In the next step, we use the mounted graphite block on the
cantilever as the tip of an atomic force microscope (AFM) in order
to transfer thin graphite samples onto a SiO$_2$/Si substrate
for subsequent device fabrication. By operating the AFM (Nanoscope IIIa
Multimode, Digital Instruments) in contact
mode with load on the graphite mounted cantilever, very thin
layers of HOPG are sheared off onto the substrate
(Fig.\ref{fig1_cleaving}(b)). This microscopic cleaving process
can be controlled by tuning the loading normal force between the
cantilever and the substrate. The van der Waals
binding energy between the graphene layers in graphite is
$\sim$~2~eV~nm$^{-2}$~\cite{Girifalco_1956}. Assuming the
graphite-substrate friction coefficient is $\sim$~1, the
required normal force to cleave off a 1~$\mu$m$^{2}$ graphite sheet from
the top of the crystallite is $\sim$~300~nN. A normal force in the
range of 10~-~2000~nN can be easily managed by commercially available
silicon cantilevers. In our experiment, by fine tuning the normal
force, square graphite crystallites with lateral size
$\sim$~2~$\mu$m and thicknesses ranging from 10~-~100~nm are
easily obtained using this process (Fig.\ref{fig1_cleaving}(c)).
Once the graphite samples are cleaved onto a substrate with
predefined alignment marks, AFM images are acquired for measuring
the height of the cleaved crystallites and registering the
position for device design. The Cr/Au (1~nm/30~-~180~nm) electrodes
are then fabricated by electron beam lithography. Typically, multiple
contacts are fabricated around the square graphite
crystallites allowing us to perform conductivity
measurements using van der Pauw method.
In addition, a degenerately doped silicon substrate
is used as a gate electrode with 500~nm of thermally grown silicon
oxide acting as the gate dielectric.

\begin{figure}
\includegraphics[width=80mm]{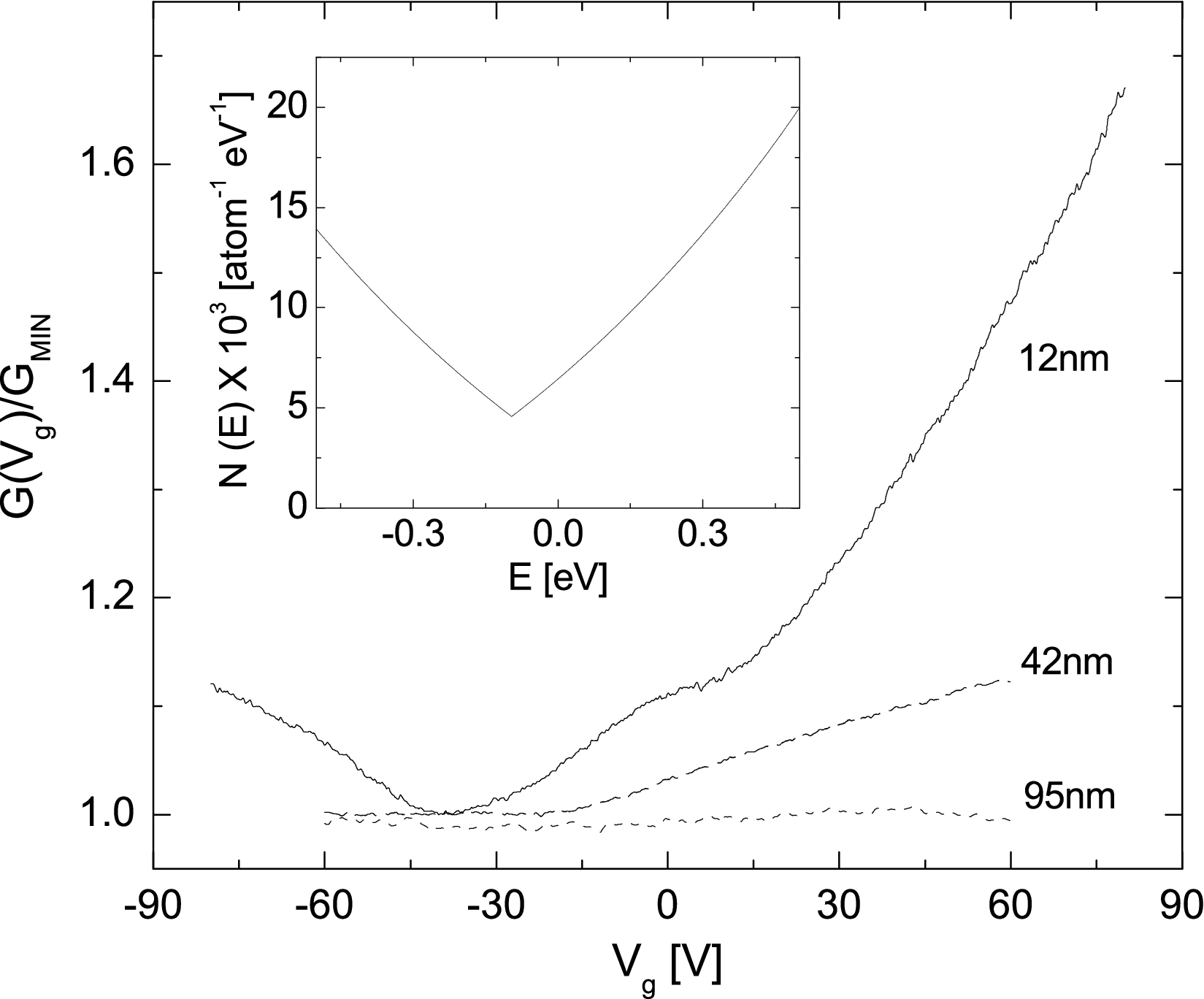}
\caption{Normalized conductance measured in samples of varying
thickness as a function of gate voltage at $T=$1.7~K.
Inset: Calculated density of states for bulk graphite \cite{McClure_1957}.}
\label{CVG}
\end{figure}

Fig.~\ref{CVG} shows the conductance, $G$, versus the gate voltage,
$V_g$, at 1.7~K for samples with varying thicknesses.
The curves are normalized with their minimum
values $G_{min}=G(V_g^{min})$, where $V_g^{min}\approx-$40~V.
As $\Delta V_g=V_g-V_g^{min}$ moves away from zero, $G(V_g)$
increases, increasing more rapidly for thinner samples.
Specifically, at $\Delta V_g=$~100~V, $G/G_{min}$ is 1.47 and 1.12
for the samples with thickness 12 and 42~nm respectively,
while no appreciable change is observed in the 95~nm sample.
This observation implies a larger electric field
effect in thinner samples, as expected.

Considering the electrostatic coupling between the samples and the
gate, the observed behavior of $G(V_g)$ at low temperatures can be
understood quantitatively as discussed below. When the gate
voltage is applied, the electrostatic potential at the surface of
the sample rises to $\phi_0$, and a charge density $n_{ind}(z)$ is
induced in the sample, where $z$ is measured from the interface of
the sample and the substrate. Following the Thomas-Fermi
approximation~\cite{A&M}, the electrostatic potential in the sample is given by
$\phi(z)=\phi_0 e^{-z/\lambda_s}$, where $\phi_0$ is a constant
(determined below) and $\lambda_s$ is the screening length.
$\lambda_s$ can be inferred from
$\phi/\lambda_s^2=-en_{ind}/\varepsilon_0$, where $e$ and
$\varepsilon_0$ are the electron charge and vacuum permittivity,
respectively. The total amount of induced charge in the
sample with thickness $d$ is related to the capacitance of the
sample to gate per unit area, $C_g$, by
$C_g(V_g-\phi_0)=-\int_0^den_{ind}(z)dz$. Evaluating this integral,
we obtain
$\phi_0=V_g/(1+C_Q/Cg)$, where the sample dependent quantum
capacitance per unit area $C_Q=\varepsilon_0(1-e^{-d/\lambda_s})/\lambda_s$.

In order to estimate the enhanced conductance in the presence of
an induced charge density gradient, we use Einstein's relation
for local conductivity $\sigma(z)=e^2N(E_F)D(E_F)$, where $N(E_F)$
and $D(E_F)$ are the density of states and and the electron diffusion
coefficient at Fermi energy $E_F$, respectively. In graphite both of
these quantities are sensitive to the change of $E_F$,
thus it should be accounted for in the determination of local
conductivity in the presence of an electric field. Assuming our
samples retain the band structure of bulk graphite, the functional
form of $N(\epsilon)$ can be obtained from graphite band structure
calculations (Fig.~\ref{CVG} inset)~\cite{McClure_1957}.
Here $\epsilon$ is measured from the
charge neutral point, where $N(\epsilon)$ has its minimum value.
For the energy range $|\epsilon|<$~300~meV,
we can approximate $N(\epsilon)\approx N_0(1+\beta |\epsilon|)$,
where $N_0=$~5.2$\times$10$^{20}$~cm$^{-3}$eV$^{-1}$ and
$\beta=$~4.8~ev$^{-1}$. A further assumption is required to estimate
$D(E_F)$ in the presence of an electrostatic potential. Here, we
adopt the simple two band (STB) model~\cite{Kelly_1981}, which
has been successful in explaining electron transport in bulk
graphite~\cite{Du&Hebard}. Assuming the scattering time, $\tau$,
and effective mass of carriers are relatively
insensitive to change of $E_F$,
the STB model yields $D(\epsilon)\propto v_F^2\tau
\propto |\epsilon+E_0|$, where $v_F$ is the Fermi
velocity and $E_0$ half the band overlap between the electron and
hole bands. With these assumptions, the normalized
conductance of the sample is given by
\begin{equation}
\label{eq_1}
G(V_g)/G_{min}=\int_0^ddzN(\epsilon)|\epsilon+E_0|/N_0E_0d
\end{equation}
where $\epsilon(z)=-e\phi_0e^{-z/\lambda_s}$. In order to compare this equation
to our experimental observations, the
values of $\lambda_s$ and $E_0$ are needed.
We use the following values from the literature:
$\lambda_s=$~0.4~nm~\cite{Visscher&Falicov} and
$E_0=$~0.015~eV~\cite{Brandt_1988}. Employing these values, we
estimate $C_g$ by comparing Eq.~\ref{eq_1} to the experimentally
observed values. The corresponding $C_g$ are 30~aF$\mu$m$^{-2}$ and
27~aF$\mu$m$^{-2}$ for 12~nm and 42~nm thick graphite devices,
respectively. These values are in reasonable agreement with the
value of $C_g$ obtained by considering the device geometry.

\begin{figure}
\includegraphics[width=80mm]{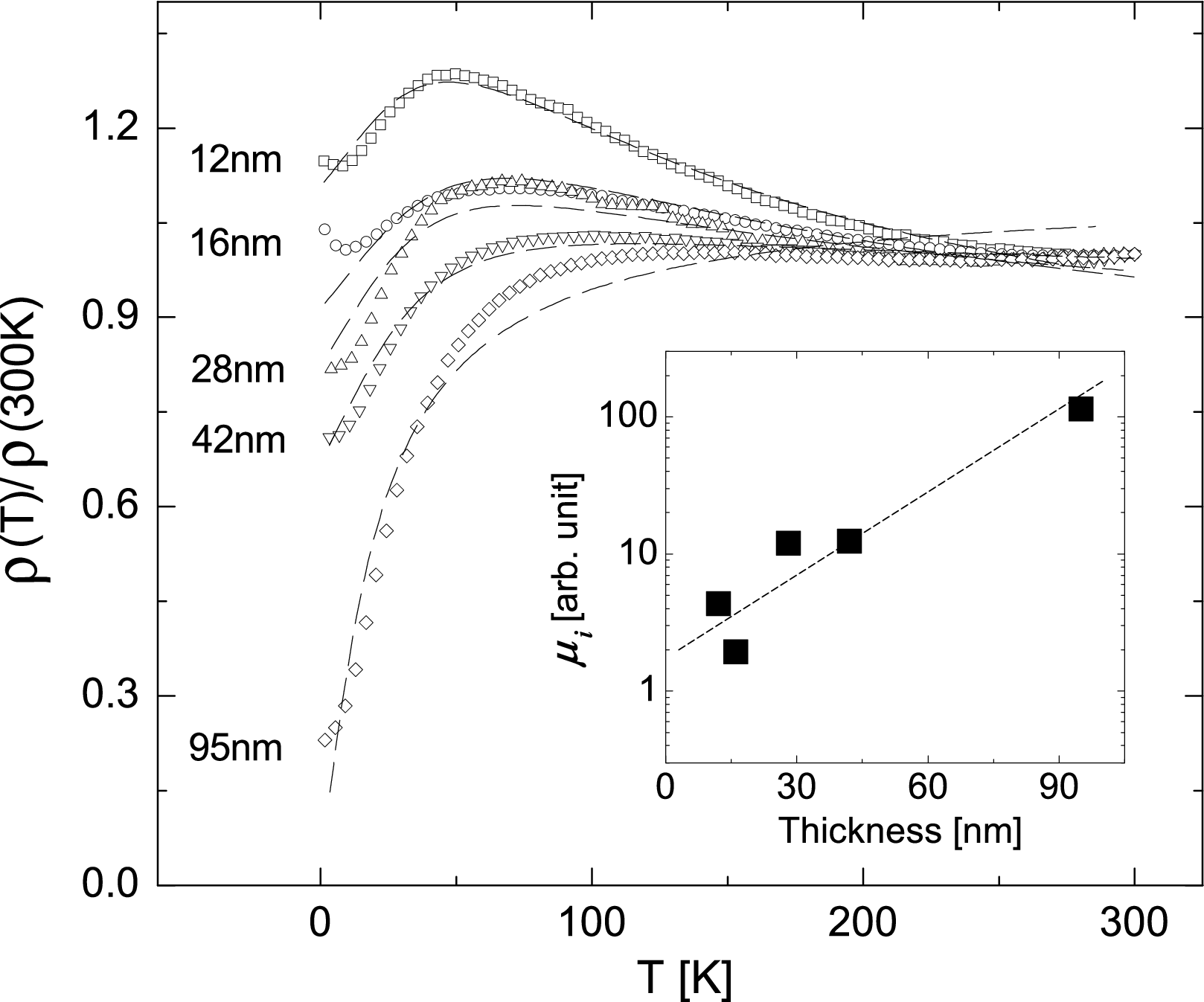}
\caption{Resistivity (normalized to its 300~K value) as a function
of temperature for different thicknesses. Broken lines are
fits with the STB model. Inset: The mobility due to
static disorder, as a result of fits to the STB model. The dotted line
is an empirical fit.} \label{RT}
\end{figure}

We now turn our attention to the temperature dependence of the
electrical conductance in samples of varying
thickness. Fig.~\ref{RT} shows the resistivity normalized
to its room temperature value,
$\rho(T)/\rho(300K)$,
as a function of temperature, $T$, for $d$ ranging from
12~-~95~nm at $V_g=$~0. While the thicker
samples show a metallic behavior ($d\rho/dT>$~0), thinner samples
develop a more complicated behavior in $T$. This behavior can be
understood by considering two competing factors as the temperature
varies. Assuming the mobility, $\mu$, is similar for the
compensating electrons
and holes in the samples, the sample resistivity is given by
$\rho=1/(|e|\mu n)$, with the
total charge carrier density $n=n_e+n_h$, where $n_e$ and $n_h$
are the carrier densities of electrons and holes, respectively. In the
STB model, for completely compensated samples (i.e.,
$\Delta V_g=$~0), $n/2=n_e=n_h=C_0k_BT\ln\big(1+\exp\frac{E_0}{k_BT})$,
where $k_B$ is Boltzmann constant and $C_0$ a constant
independent of $T$~\cite{Kelly_1981}.
We further assume
that the mobility can be expressed as $\mu^{-1} = \mu_i^{-1}
+ A_0T$, where $\mu_i^{-1}$ is the contribution from static
scattering centers and $A_0$ is a constant depending on
electron-phonon scattering in graphite. Then, as $T$
increases, $\mu$ decreases due to increasing electron-phonon
scattering, while $n$ increases rapidly due to the thermal excitation of
charge carriers around $E_F$. Using $E_0$ and $A_0\mu_i$ as fitting
parameters, this model fits well with our experimental
observations over a large temperature range and samples with
varying thicknesses (broken lines in Fig.~\ref{RT}).
As a result of the fit, we obtain $E_0\sim$~10~-~20~meV, which is in
reasonable agreement with the values found in the literature~\cite{Brandt_1988}.
In addition, we obtain $A_0\mu_i$ from the same fit. Assuming
$A_0$ is similar for all measured samples, the $\mu_i$ for different
samples can be compared. The inset of Fig.~\ref{RT} shows
decreasing $\mu_i$ as sample thickness is reduced.
An increasing contribution of boundary scattering at
the sample surfaces can explain this behavior. We note, however, that
the reduction of $\mu_i$ seems to saturate as $d\rightarrow0$.
In addition, our recent galvanomagnetic
measurements~\cite{Zhang_PRL} in these samples showed that the
quantum oscillations in electron transport can be readily observed
at high magnetic fields. Both observations suggest that
the quality of the sample has not degraded after our micromechanical
extraction from the bulk.

We thank Y.~Wu for help in sample processing, V.~Oganesyan
and P.~McEuen for
helpful discussions. This work is supported primarily by the
Nanoscale Science and Engineering Initiative of the National
Science Foundation under NSF Award Number \mbox{CHE-0117752} and
by the New York State Office of Science, Technology, and Academic
Research \mbox{(NYSTAR)}.

\end{document}